\pdfoutput=1 
\documentclass[12pt]{article}

\usepackage{inputenc}

\usepackage{amsmath,amssymb,bbm,color}
\usepackage{amsbsy}
\usepackage{euscript}
\usepackage{graphicx}
\usepackage{hyperref}
\usepackage{cite}
\usepackage{enumerate}
\numberwithin{equation}{section}

\usepackage[]{changes} 

\newcommand{\order}[1]{${\cal O}(#1)$}

\newcommand{\kksem}{{\tt  KKsem}}
\newcommand{\kkmc}{{\tt   KKMC}}
\newcommand{\koralz}{{\tt KORALZ}}
\newcommand{\zfitter}{{\tt ZFITTER}}

\def\Order#1{${\cal O}(#1)$}

\begin{document}

\begin{titlepage}

\begin{flushright}
\bf IFJPAN-IV-2019-12
\vskip 1pt
\bf IRFU-19-18
\end{flushright}

\vspace{5mm}
\begin{center}
    {\Large\bf Precision measurement of the Z boson 
    to electron neutrino coupling
    at the future circular colliders$^{\star}$ }
\end{center}

\vskip 5mm
\begin{center}
{\large R.\ Aleksan$^{a}$ and S.\ Jadach$^{b}$}

\vskip 2mm
\vspace{1mm}
{\em $^a$IRFU, CEA, Universit\'e Paris-Saclay, 91191 Gif-sur-Yvette cedex, France\\ }
{\em $^b$Institute of Nuclear Physics, Polish Academy of Sciences,\\
  ul.\ Radzikowskiego 152, 31-342 Krak\'ow, Poland}\\
\end{center}

\vspace{2mm}
\begin{abstract}
\noindent
At the high luminosity electron-positron circular colliders
like FCC-ee in CERN and CEPC in China 
it will possible to measure very precisely
$e^+e^-\to Z\gamma$ process with subsequent Z decay into 
particles invisible in the detector, 
that is into three neutrina of the Standard Model
and possibly into other weakly coupled neutral particles.
Apart from the measurement of the total invisible width
(which is not the main subject of this work)
this process may be used as a source of $Z$ coupling
to electron neutrino -- known very poorly.
This is possible
due to the presence of the $t$-channel $W$ exchange
in the $e^+e^-\to \nu_e \bar\nu_e \gamma$ channel
which deforms slightly spectrum of the photon.
We are going to show experimental investigation of this
effect, for $\sim 10$ inverse atobarn accumulated
luminosity, which can provide measurement of the $Z-\nu_e$
coupling with statistical error of order 1\%.
The estimation of the systematic experimental error will require more work,
but most likely it will be of similar size.
\end{abstract}

\vspace{50mm}
\footnoterule
\noindent
{\footnotesize
$^{\star}$This work is partly supported by
 the Polish National Science Center grant 2016/23/B/ST2/03927
 and the CERN FCC Design Study Programme.
}

\end{titlepage}


\newpage
\section{Introduction}

\noindent Recent measurements in the B meson sector have shown evidences for 
lepton universality violation at the 3 to 4 standard deviations ($\sigma$) levels. 
Without being exhaustive, 
deviations from the Standard Model (SM) are observed by comparing semileptonic 
B decays $B\to D^{(\ast )} \tau \nu$  on the one hand, 
and $B\to D^{(\ast )} \ell \nu$ with $\ell = \mu , e$ 
on the other hand~\cite{babar:0,babar:1,belle:1,lhcb:0,lhcb:1}.
Discrepancies are also observed in comparing the decays $B\to K^{(\ast )} \mu \nu$  
and $B\to K^{(\ast )} e \nu$~\cite{belle:2,babar:2,lhcb:2}. 
In contrast, lepton universality is verified 
at the $1\%$ and 0.1$\%$ level in $W$ and $Z$ decays respectively, 
involving charged leptons~\cite{pdg:1}, although there is a slight tension 
at the $3\sigma$ level in the W decay $W\to \tau \nu$ versus the average  
$W\to \ell \nu \ ( \ell = \mu,e)$. 
In the same vein,  one notes that neutrino counting in $Z$ decays shows 
a very slight deficit ($2\sigma$), 
with $N_\nu =2.984\pm 0.008$~\cite{pdg:1}.
(Although, corrections due to beam-beam effects 
 of recent Ref.~\cite{Voutsinas:2019hwu}
 change this result to $N_\nu =2.992\pm 0.008$.)
Therefore, to shed further light in this area and complement these tests, 
the precision of which should be considerably improved 
in the near and long term future at HL-LHC, SuperKEKB and FCC, 
it might be very useful to study universality in $Z\to \nu_\ell\bar{\nu}_\ell$ decays. 
In general, this is a non trivial task since neutrinos are difficult to detect 
and thus is the identification of their species. 
In this paper we investigate a method to achieve this objective 
by ``making the neutrino flavors visible in Z decays''.
The future high-energy circular electron-positron collider
FCC-ee~\cite{Gomez-Ceballos:2013zzn,Mangano:2018mur,Benedikt:2018qee} 
would be an ideal tool for this purpose.

\section{Testing Universality of neutrinos in $Z^0$ decays}

The charged and neutral current interaction of neutrino have been observed since 
a long time in fixed target experiments using muon neutrino beams. 
Although the production of a charged lepton in the final state tags the incoming neutrino flavor 
in charged current interaction for Deep Inelastic Scattering (DIS) events, 
the same procedure cannot be used in neutral current since the final state neutrino is not detected.
The flavor is only inferred indirectly by comparing the neutral current events 
to the SM expectations.
More recently, the SNO experiment has measured  the charged and neutral current interactions 
of solar electron neutrinos.  
The comparison of the respective rates has shown that different types of neutrinos 
are involved in neutral current events~\cite{sno:1} than the one ($\nu_e$) 
observed in charged current. 
But here again, the flavor of the neutrinos interacting in the neutral 
current events is not identified and thus any extraction of $Z\nu\nu$ coupling is theory dependent. 
Furthermore the uncertainty on the incoming neutrino flux does not enable very precise measurements.
The overall situation is summarized 
by the Particle Data Group~\cite{pdg:1} for the $Z$ couplings 
to $\nu_e$ and $\nu_\mu$ as follows:
\begin{equation}
\begin{split}
g_{Z}^{\nu_e}  = & 1.06 \pm 0.18   \\
g_{Z}^{\nu_\mu}=  & 1.004 \pm 0.034 
\end{split}
\label{eq:bw_Z}
\end{equation}
In the following we are going to present a method to measure
the individual $Z\nu_e\nu_e$ coupling, which is so far poorly measured.

\section{The method}

The combined strength of couplings of $Z$ boson to all particles contributing
to $Z$ invisible width was at LEP experiments conveniently parametrized~\cite{ALEPH:2005ab}
in terms the so called neutrino number parameter $N_\nu$, equal 3 in the SM.

\begin{figure}[!t]
  \centering
  \includegraphics[width=0.9\textwidth]{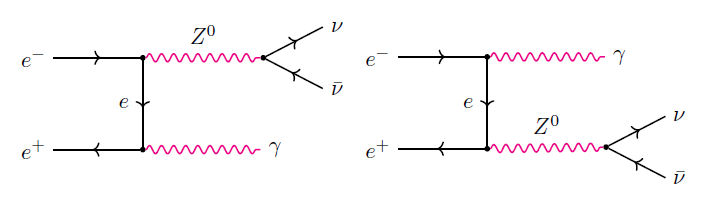}
  \caption{\label{fig:diagram-Zgamma} Production of flavor-untagged $\nu$ through the process $e^+e^-\to Z^0\gamma \to \nu\bar{\nu}\gamma$.
  }
\end{figure}

Two sensitive ways to measure $N_\nu$ have been used in the past:


\begin{enumerate}
\item 
Precise determination of the $Z^0$ production peak in $e^+e^-$ collisions 
and the $Z$ line shape. This is the most precise method 
with $N_\nu =2.984\pm 0.008$~\cite{ALEPH:2005ab,pdg:1}, 
thanks to the statistics and the precise beam energy determination at LEP.
\item 
Identification and counting of the $Z$ boson radiative return (ZRR) process
in a center of mass energy above the $Z$-pole, 
i.e. using Initial State Radiation (ISR) $e^+e^-\to \gamma X$,
see diagrams in Figure~\ref{fig:diagram-Zgamma}.
LEP has measured $N_\nu =2.92\pm 0.05$~\cite{Abbiendi:2000hh,pdg:1}.
\end{enumerate}

\begin{figure}[!t]
  \centering
  \includegraphics[width=0.9\textwidth]{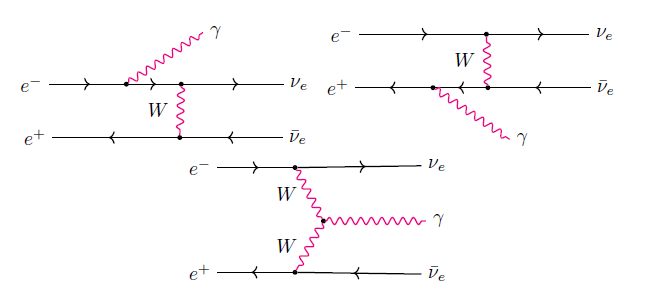}
  \caption{\label{fig:diagram-Wgamma}
  Production of flavor-tagged $\nu_e$ through the process 
  $e^+e^-\to \nu_e\bar{\nu_e}\gamma$ with W exchange.
  }
\end{figure}

\noindent
However in both cases, 
one has only derived the total number of light neutrino species ($N_\nu$) assuming universality of the $Z$ coupling to the neutrino species, 
i.e. without measuring the individual couplings. 
Indeed, by determining the $Z$ production peak cross section and the $Z$ width, 
one deduces its invisible cross section.
\begin{equation}
\begin{split}
&\sigma(e^+e^-\to Z \to \mathrm{invisible}) =  \\
& (g_{Z}^{\nu_e} \mathcal{A}_{Z}^{\nu_e})^2 
 + (g_{Z}^{\nu_\mu}\mathcal{A}_{Z}^{\nu_\mu})^2 
 + (g_{Z}^{\nu_\tau}\mathcal{A}_{Z}^{\nu_\tau})^2 
 + (g_{Z}^{X}\mathcal{A}_{Z}^X)^2,
\end{split}
\end{equation}
where $g_{Z}^{\nu}$ and $ \mathcal{A}_{Z}^{\nu}$ are the individual couplings of Z to neutrinos 
and the well know Breit-Wigner amplitudes  respectively. Similarly $g_{Z}^{X}$ 
and $\mathcal{A}_{Z}^X$ are related to Z decays to invisible new physics, 
which couples to $Z$, if any. 
Since all  $ \mathcal{A}_{Z}$ are identical for fermions, one gets
\begin{equation}
N_\nu \sim
(g_{Z}^{\nu_e})^2 + (g_{Z}^{\nu_\mu})^2 + (g_{Z}^{\nu_\tau})^2 + (g_{Z}^{X})^2
\label{eq:N_nu}
\end{equation}
Normalizing SM couplings to one, $ g_{Z}^{\nu_e} =g_{Z}^{\nu_\mu}=g_{Z}^{\nu_\tau}=1 $
and $g_{Z}^{X}=0$, one obtains  $N_\nu =3$.

At FCC-ee~\cite{Gomez-Ceballos:2013zzn,Mangano:2018mur,Benedikt:2018qee}, 
a very significant  improvement (by several orders of magnitude) 
is expected for the determination of $N_\nu$, 
due to high luminosities achievable around the $Z$-pole and at higher energy
and reduction of the experimental and theory uncertainties~\cite{Jadach:2019bye}.

Obviously, there is no means of discriminating one coupling constant from another 
in the process of Figure~\ref{fig:diagram-Zgamma}, since only the sum of the couplings is measured. 
For the following in this paper, 
let us define the parameter $\eta$ rescaling $Z$ couplings as follows
\begin{equation}
g_{Z}^{\nu_e}     =  \sqrt{1+\eta},   \quad
g_{Z}^{\nu_\mu}   =   1,  \quad
g_{Z}^{\nu_\tau}  =  \sqrt{1-\eta},
\label{eq:gZ}
\end{equation}
where the deviation of $g^{\nu_e}_Z$ from the SM is compensated
by the opposite deviation of $g^{\nu_\tau}_Z$, while keeping constant
more precisely measured total invisible $Z$ width and $g^{\nu_\mu}_Z$.

\begin{figure}[!t]
  \centering
  \includegraphics[width=165mm]{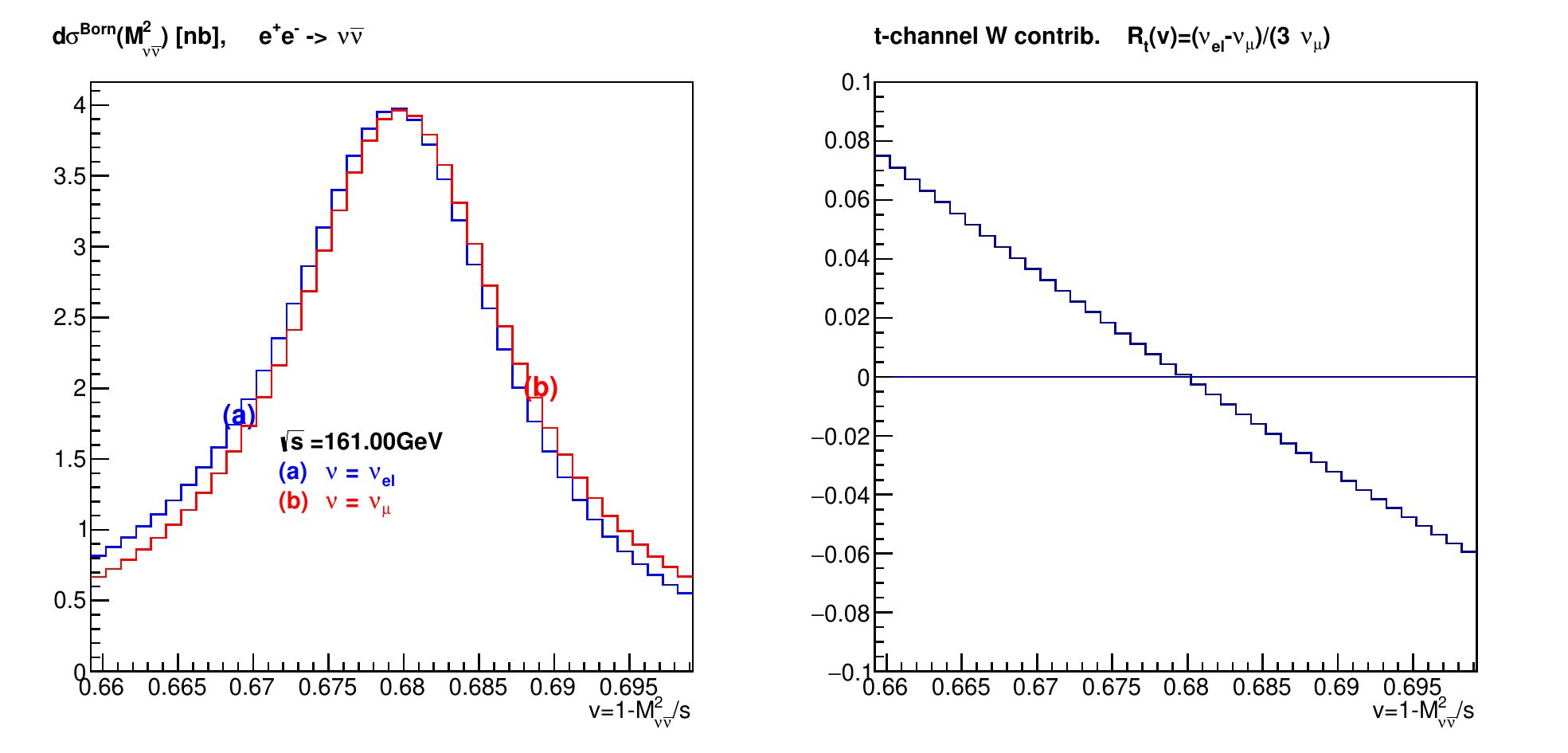}
  \caption{\sf
  The LHS plot shows $\sigma^{Born}(M^2_{\nu\bar\nu})$  for
  $ e^+e^- \to \nu\bar\nu, \nu= \nu_e, \nu_\mu $
  as function of variable $v=1-M^2_{\nu\bar\nu}/s$,
  for $s^{1/2}=161$GeV,
  the same as in the following MC results for ZRR process.
  The RHS plot illustrates relative contribution of the $t$-channel $W$-exchange
  diagram in the Born cross section for all three neutrinos.
  }
  \label{Fig:BornNu}
\end{figure}

The important point is that in the $e^+e^- \to \nu_e \bar\nu_e\gamma$ process
there are additional diagrams producing electron neutrinos in the final state,
with $t$-channel $W_t$ boson exchange,
see Figure~\ref{fig:diagram-Wgamma}. 
These diagrams interfere with the ones in Figure~\ref{fig:diagram-Zgamma}.
Therefore, observing this interference in the $\gamma$ energy spectrum
related to the $\nu\bar\nu$ invariant mass%
\footnote{%
 Here and in the following $E_\gamma$ is total energy of one or more photons detected above
 certain minimum energy and minimum angle from the beams in the ZRR process. 
 Should there be multiple $\gamma$, an additional term proportional to $M_{n\gamma}^2$ is present but can be safely neglected as it is very small.
},
\begin{equation}
M_{\nu\bar\nu}^2  
\simeq   s - 2\sqrt{s} E_{\gamma}\quad  \mathrm{or}\quad  
v= \frac{E_{\gamma}}{E_{beam}}  \simeq  1 - \frac{M_{\nu\bar\nu}^2}{s},
\label{eq:m_recoil}
\end{equation}
would lead to the measurement of the $Z-\nu_e$ coupling ($g_Z^{\nu_e}$),
as aimed in this paper.
We shall refer in the following to this interference in short as 
{\em $Z_s\otimes W_t$ interference}.

In the left part of Fig~\ref{Fig:BornNu} we show how big is the effect
of the $t$-channel $W$-exchange contribution of Figure~\ref{fig:diagram-Wgamma}
comparing Born cross section $\sigma^{Born}(M_{\nu\bar\nu}^2)$
for $\nu=\nu_e$ and $\nu=\nu_\mu$  near the $Z$ peak,
that is in the range $|v-v_Z|\leq 0.02$, $v_Z=1-M_Z^2/s$,
which translates into $88.299 \geq M_{\nu\bar\nu}\geq 93.987$
for  $s^{1/2}=161$GeV.
The relative effect of the $W$-exchange with respect to all 3 neutrino case
is up to 8\%, changing sign in the middle of the $Z$ peak.
Born cross section is calculated using expressions of eqs.~(2.9-2.11)
in Ref.~\cite{Bardin:2001vt}.
We have checked that we reproduce benchmark Table~1 
of Born cross sections and forward-backward asymmetries%
\footnote{In Ref.~\cite{Bardin:2001vt} 
  it was obtained in using non-MC programs \kksem\ and \zfitter.
}
in this paper.

From Figure~\ref{Fig:BornNu} it is obvious that in the ZRR process our aim will
be to measure asymmetric deformation of the $Z$ resonance shape in
the photon spectrum, that is its {\em skewness}.
Of course, QED effects will also contribute to the skewness 
of the $Z$ lineshape ZRR spectrum, hence very good quality MC program for the ZRR
of the \kkmc\ class~\cite{Jadach:1999vf}, or even better, 
will be indispensable to sort out QED effects
in the FCC-ee data analysis.

The contribution of the $W$-exchange diagram near the Z-peak is almost entirely
due to the interference of the dominant s-channel $Z$-resonant amplitude 
(which contains precious coupling of the $Z$ to electron neutrino)
and the trivial $W$-exchange diagram,
as seen clearly from the presence of the zero in the middle of the $Z$ peak --
while $W$-exchange diagram squared is negligible.

The diagram contribution pattern is quite different
for ZRR at low $v$ and high $M_{\nu\bar\nu}$, where $W$-exchange diagram
dominates over the $Z$-resonant diagram by order of magnitude%
\footnote{For example in the range $v\in (0.2,0.4)$
   corresponding to $M_{\nu\bar\nu}\in (125,144)$GeV at  $s^{1/2}=161$GeV.
}.
Again, it is the interference of the two, which could provide valuable information
on the $Z$ couplings to electron neutrino.
Due to much smaller cross sections this option looks less attractive,
nevertheless it requires quantitative study in the future.

The issue of how the invisible width of $Z$ boson 
parametrized in terms of $N_\nu$ could indicate New Physics
was elaborated in many papers, see for instance Refs.~\cite{Carena:2003aj,Ellis:2019zex}.
The present work is the first dedicated study 
on how to extract $Z$ coupling to electron neutrino
taking advantage of extraordinary luminosity at FCC-ee.
The question how this measurement 
could influence searches of New Physics deserves separate studies.

For completeness, let us finally note that a very different 
and more straight forward way for identifying the neutrino species produced in Z decays 
would be to observe the interaction of the neutrinos within the FCC-ee detector. 
Indeed, with an integrated luminosity of 150 $ab^{-1}$ at the $Z$-pole, 
some $2.4 \times 10^{12}$ neutrinos are produced. 
Although this figure is large, 
unfortunately the charged current neutrino cross section with 
$E_\nu = 45$\ GeV is low; $\sim 0.3 \ pb$. 
Assuming a tracking area with 1 radiation length (which is far more than usual trackers) 
only $\sim 3$ interactions are expected. 
So this direct detection method seems unpractical, 
unless one develops a dedicated segmented detector with some 100 $X_0$.
We expect similar size effect in the spectrum of the ZRR photon.

\section{Matrix element of KKMC}

In the following numerical studies we shall use a version on \kkmc\ , 
which features matrix element of the $e^+e^-\to \nu\bar{\nu} +n\gamma$
present in \kkmc\ since version 4.19.
The source code of the version 4.19 \kkmc\ is available from
\href{http://jadach.web.cern.ch/jadach/KKindex.html}{http://jadach.web.cern.ch/jadach/KKindex.html}\\
or
\href{http://192.245.169.66:8000/FCCeeMC/wiki/kkmc}{http://192.245.169.66:8000/FCCeeMC/wiki/kkmc}.

The original 1999 version of \kkmc\ of ref.~\cite{Jadach:1999vf} 
did not yet include a good quality
matrix element for the neutrino pair production process.
This is why during 1999/2000 LEP Physics Workshop~\cite{Kobel:2000aw},
theoretical studies of the $\nu\bar{\nu}\gamma+ n\gamma$ final states were based on
{\tt KORALZ}~\cite{Jadach:1993yv},
{\tt NUNUGPV}~\cite{Montagna:1996ec,Montagna:1998ce} and
{\tt GRC4F}~\cite{Fujimoto:1996wj,Kurihara:1999vc} MC programs.
The conclusion was at the time that predictions of these program can be trusted to within 2-3\%.
\koralz\ has featured approximate matrix element for two real photons and
approximate matrix elements for $t$ channel $W$ exchange (in case of $\nu_e$),
see also Ref.~\cite{Colas:1990ef}.
On the other hand {\tt NUNUGPV} and {\tt GRC4F}, 
have included exact matrix element for two photons,
but soft photon resummation was
implemented in {\tt GRC4F} through a QED parton shower 
and in {\tt NUNUGPV} through electron structure function formalism, 
instead of coherent exclusive exponentiation of ref.~\cite{Jadach:1999vf}.
Both codes adopted methods to remove the double counting of radiation
between matrix element and resummation.
%
The virtual corrections to $\nu\bar{\nu}\gamma$ process were known from
earlier work of Ref.\cite{Igarashi:1986ht} and used in some of these programs%
\footnote{Virtual corrections to $W$-exchange diagram were neglected in this work.}.

Before the end of LEP era matrix element of \kkmc\ 
for neutrino pair production process was upgraded
and documented in a fine detail in Ref.~\cite{Bardin:2001vt}.
In particular $W$-exchange diagram for electron neutrino channel was implemented
and the electroweak (EW) library DIZET, 
the same as in \zfitter~\cite{Bardin:1999yd}, was added.
The validity of the implementation of EW corrections in \kkmc\
was cross-checked in this work by means of direct comparison
of \zfitter\ and \kkmc\ for the $e^+e^-\to\nu\bar{\nu}$ process%
\footnote{
  For the $e^+e^-\to  \nu\bar{\nu}\gamma$ process EW corrections in \kkmc\ 
  enter in the soft photon approximation thanks to soft photon resummation.}
in spite that it is not seen experimentally.

Later on, in Ref.~\cite{Was:2004ig},
the exact matrix element for $e^+e^-\to \nu\bar{\nu} +2\gamma$ process
(as implemented in \kkmc) was also analysed in a great detail 
focusing on the delicate issue of the QED gauge invariance,
especially in case of photon emissions out of the $W$ boson
exchanged in the $t$-channel.

The \kkmc\ program is a general purpose MC event generator for producing
pair of any kind of charged lepton  (except electron), neutrino or quark.
The \Order{\alpha^0} Born in \kkmc\ is the $e^+e^-\to f\bar{f}$  process,
completed with the \Order{\alpha^1}, \Order{\alpha^2} QED corrections
and \Order{\alpha^1} EW corrections.
The case of neutrino is special because the \Order{\alpha^0} Born process
$e^+e^-\to \nu\bar{\nu}$ is not visible in the detector, 
hence it is $e^+e^-\to f\bar{f}\gamma$ which can be treated as \Order{\alpha^0} Born process.
(This kind of convention is used in {\tt NUNUGPV} and {\tt GRC4F}.)

In \koralz\ and \kkmc\ it is  the $e^+e^-\to \nu\bar{\nu}$ process being the Born process,
while $e^+e^-\to f\bar{f}\gamma$ is regarded as \Order{\alpha^1}.
Here and in the following we adopt the above convention.
Matrix element of \kkmc\ features the complete \Order{\alpha^2} QED corrections
(instead of the \Order{\alpha^1} in case of Born being $e^+e^-\to f\bar{f}\gamma$).
It means that \kkmc\ includes exact matrix element 
for $e^+e^-\to f\bar{f} +2\gamma$~\cite{Was:2004ig}
and complete virtual corrections to $e^+e^-\to f\bar{f}\gamma$.
Strictly speaking, as explained in Ref.~\cite{Bardin:2001vt},
one-loop corrections to $W$-exchange contribution 
for the electron neutrino pair process in \kkmc\
are taken in certain low energy approximation 
and will have to be improved in the future.

At the practical level, there is no possibility in \kkmc\ to request through input parameters
that at least one ZRR photon visible above certain minimum
energy and angle (with respect to beams) is always present%
\footnote{This is due use of the basic MC algorithm for generating multiphoton events, 
  the same as for other final fermions.}
in the generated MC sample.
One has to generate photons all over the entire phase space and then select
a subsample of the events with one or more ZRR photons.
This costs  factor $\sim 10$ loss in terms of the CPU time
for typical event selection of the ZRR process.

\begin{figure}[!t]
  \centering
  \includegraphics[width=165mm]{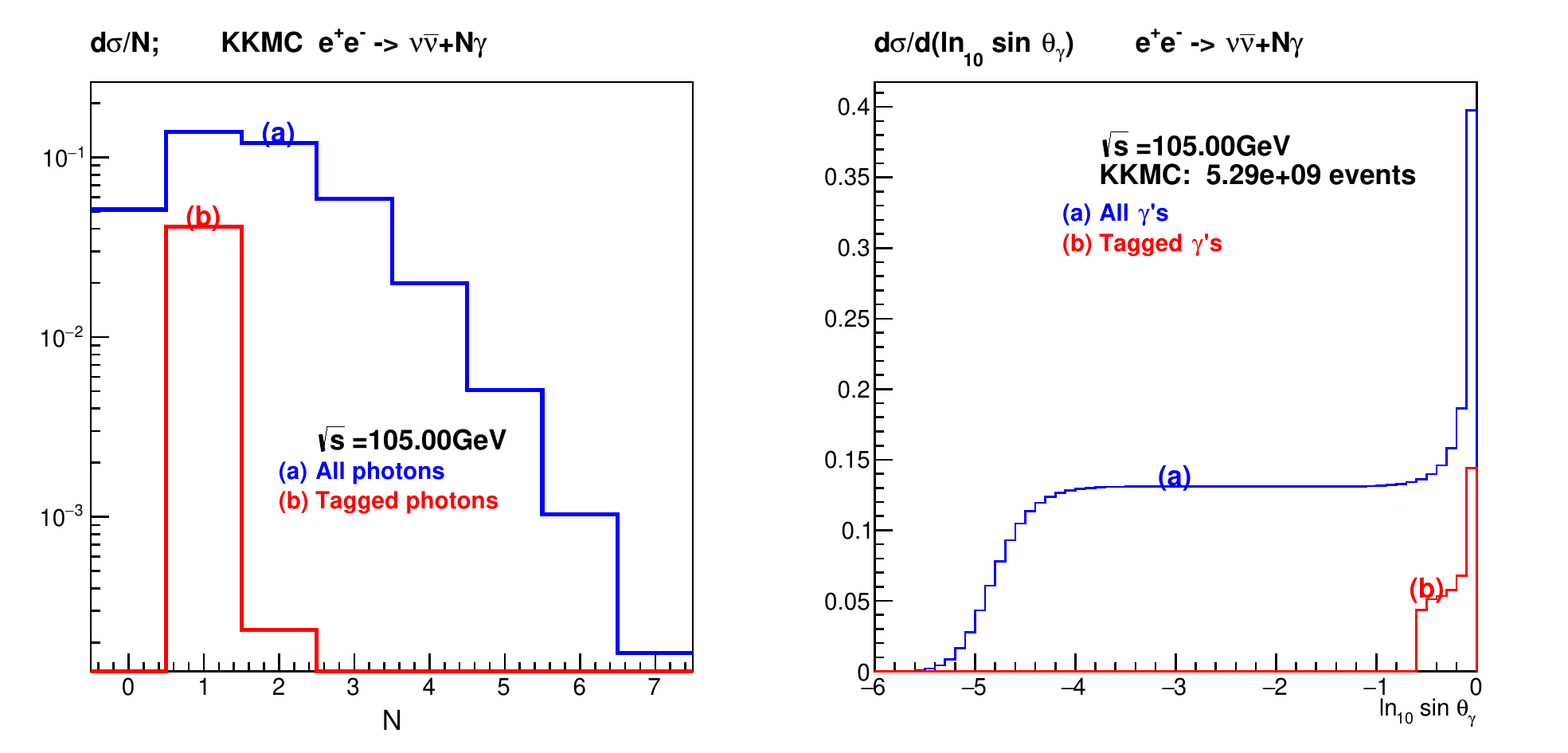}
  \caption{\sf
  The distributions of photon multiplicity and photon angle in the process
  $ e^+e^- \to \nu\bar\nu(N\gamma),\; \nu=\nu_e,\nu_\mu,\nu_\tau,\;$
  (a) without restricting photons and (b) for ZRR subsample.
  }
  \label{Fig:NPhot}
\end{figure}

\begin{figure}[!t]
  \centering
  \includegraphics[width=75mm]{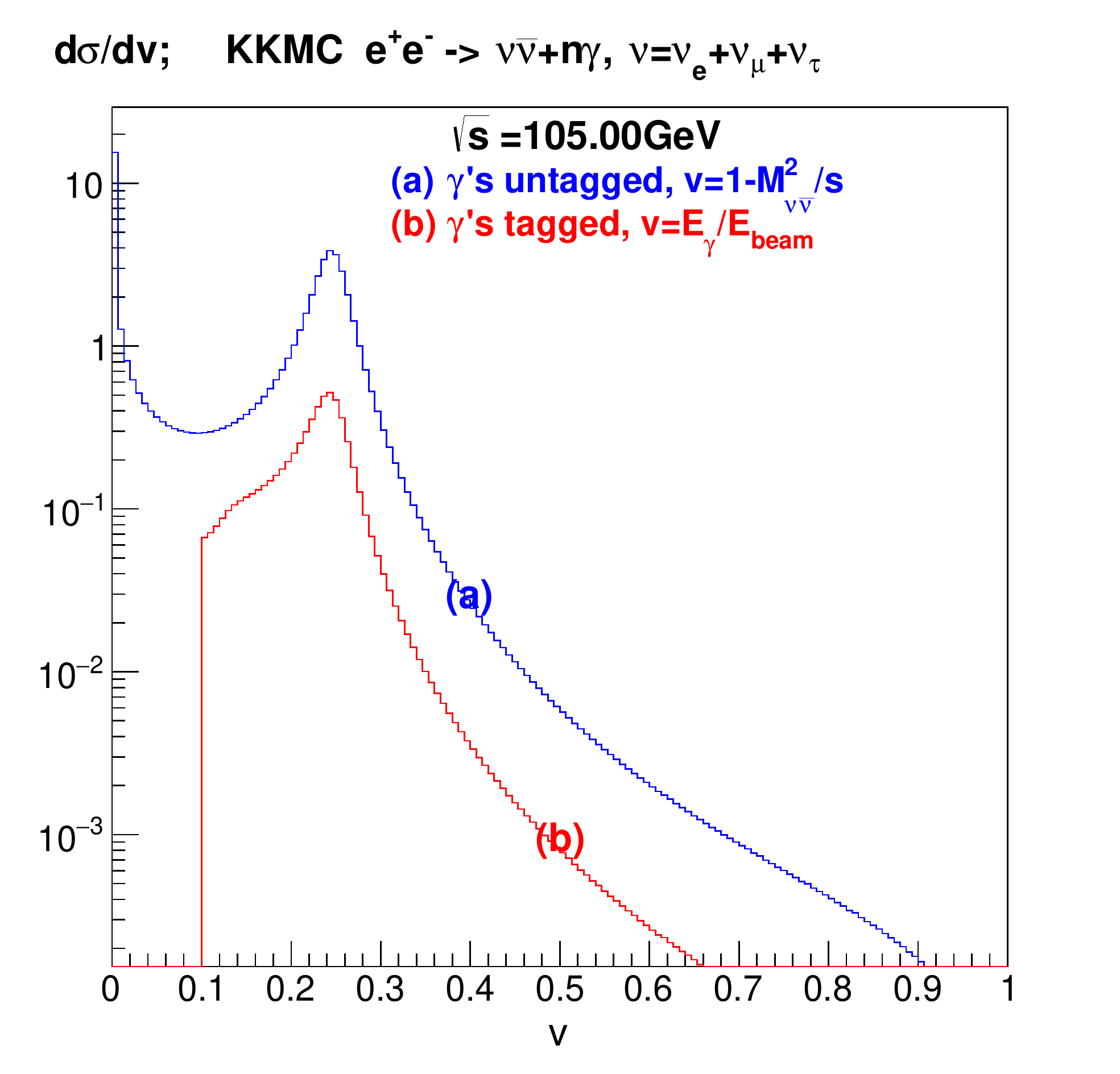}  
  \includegraphics[width=75mm]{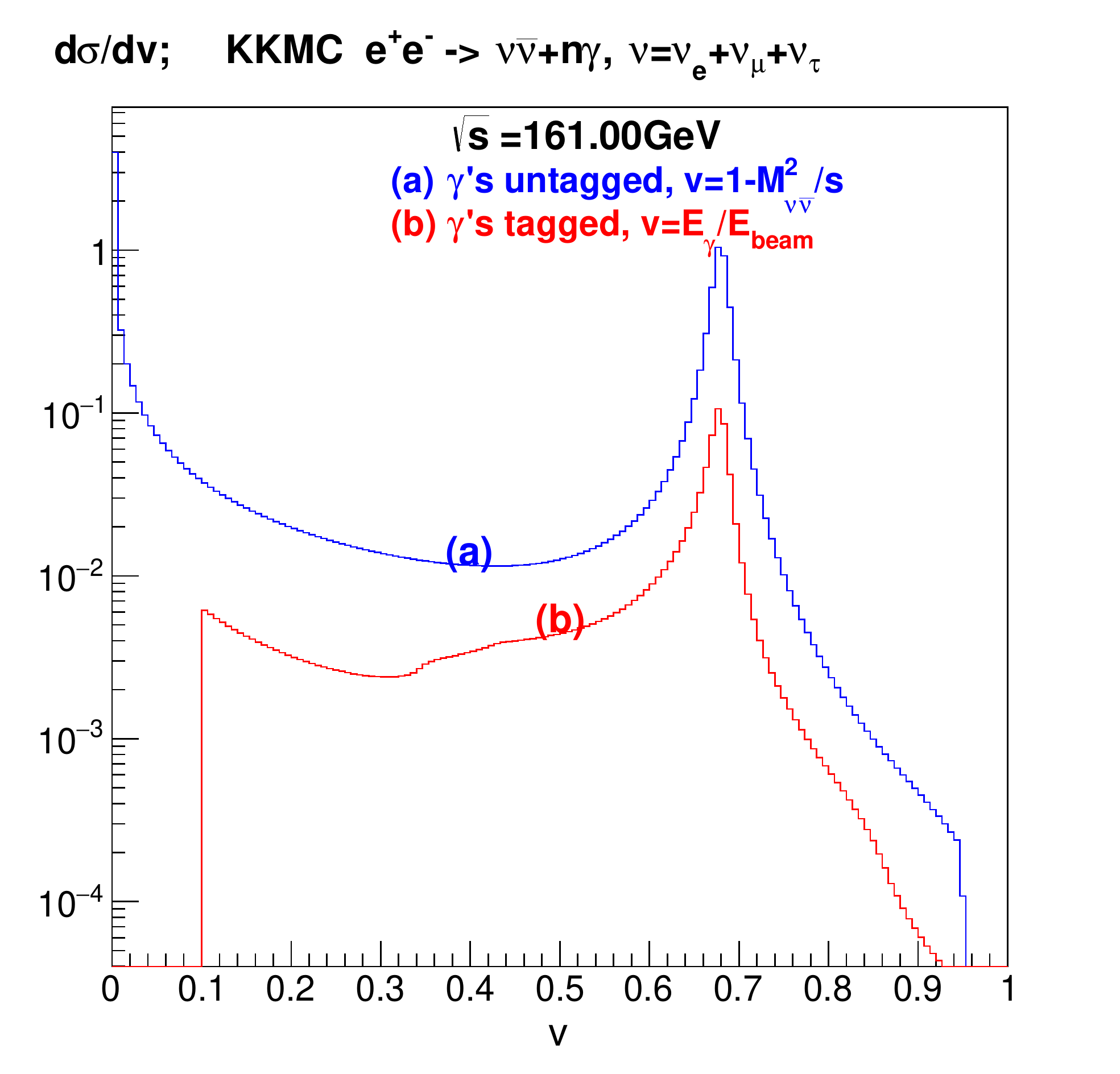}
  \caption{\sf
  The distributions of photon energy in the processes
  $ e^+e^- \to \nu\bar\nu(+n\gamma),\; \nu=\nu_e+\nu_\mu+\nu_\tau $
  (a) without restricting photons and (b) for ZRR subsample.
  }
  \label{Fig:VPhot}
\end{figure}

\section{Monte Carlo results}

Monte Carlo results in this Section were obtained in two \kkmc\ runs
at $105$GeV and $161$GeV with statistics of $\sim 4\cdot 10^9$ weighted events.
Figures~\ref{Fig:NPhot} and \ref{Fig:VPhot},
illustrate general features of the ZRR process.

Our event selection criteria for photons in the ZRR process include requirement
of sum of the photon energies being above $0.10 E_{beam}$,
each photon angle with respect to incoming beam to be above $15^{\circ}$,
and each photon transverse momentum being above $0.02 E_{beam}$.

In the LHS plot of Figures~\ref{Fig:NPhot} we see photon number distribution
for all photons generated by \kkmc\ for the $e^+e^-\to \nu\bar\nu +n\gamma$ process
above IR cut-off $10^{-5}E_{beam}$
and for photons which pass the above criteria of the ZRR process.
As we see, only $\sim 1\%$ fraction of ZRR events has two photons.

The RHS plot of Figure~\ref{Fig:NPhot} presents the same two classes of MC events
in the log of photon angle with respect to beams.
The natural cut-off at $\theta \sim m_e/\sqrt{s}$ due to small but finite electron
mass is clearly seen, and the ZRR cut-off at $15^{\circ}$ is seen as well.
The rejection rate from all MC events down to ZRR class is of order $\sim 10$.

For completeness in Figure~\ref{Fig:VPhot}
we also show the entire distribution of the energy of all ISR photons
and for visible photons in the ZRR MC sample
for two energies, 105GeV and 161GeV.

\begin{figure}[!t]
  \centering
  \includegraphics[width=160mm]{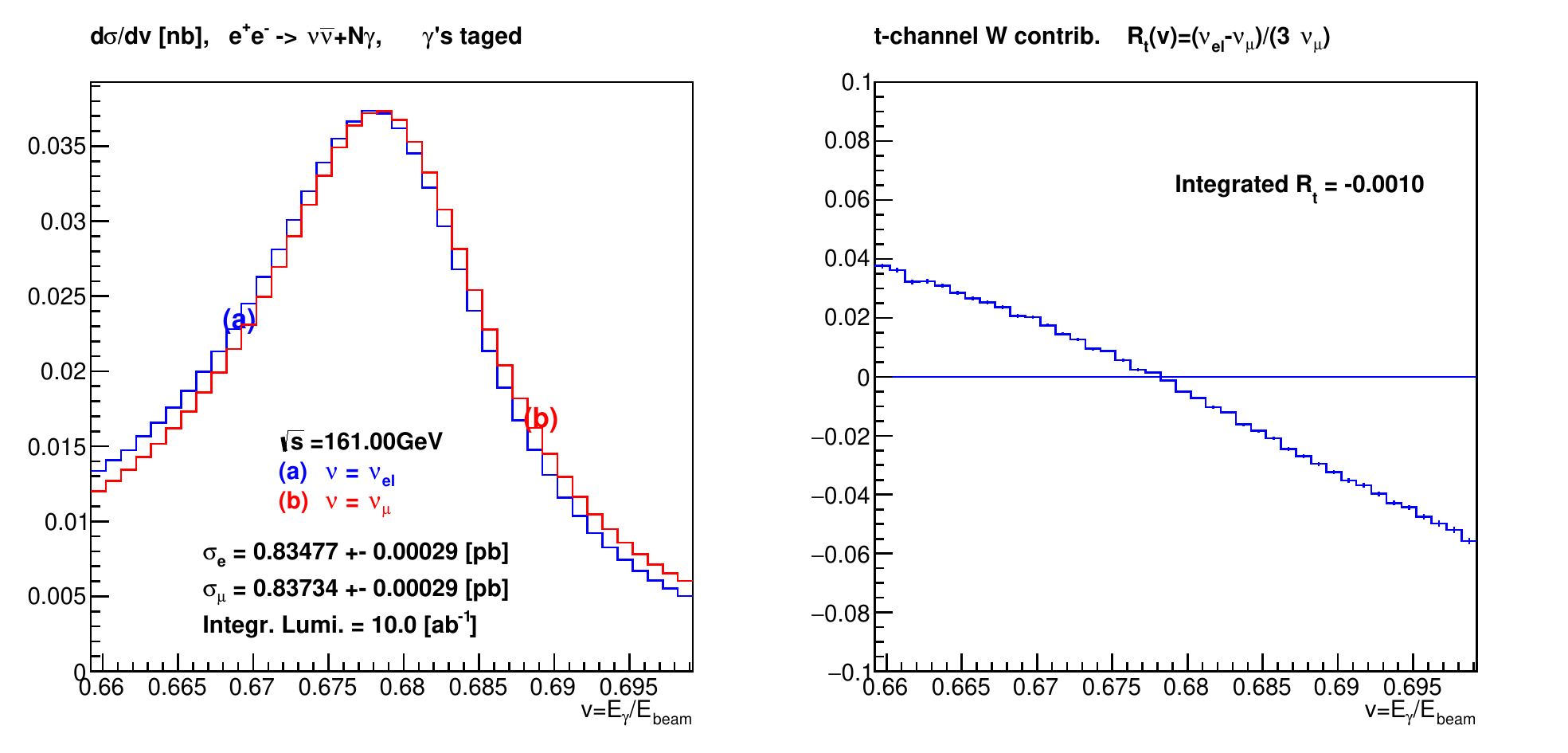}
  \includegraphics[width=160mm]{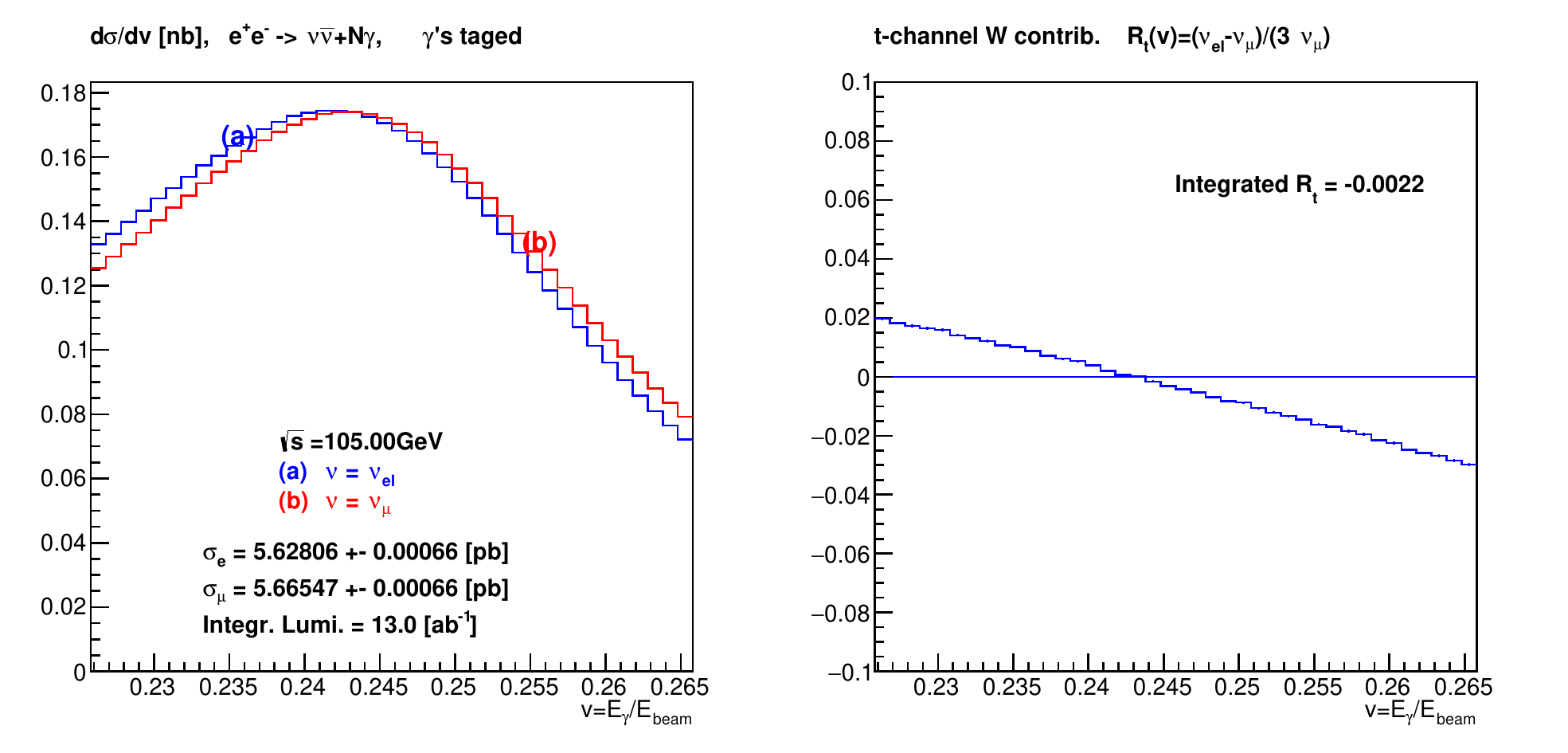}
  \caption{\sf
  Examining difference of the ZRR photon spectrum 
  for electron neutrino and muon neutrino channels
  due to $t$-channel W exchange.
  }
  \label{Fig:NuDif2}
\end{figure}

\begin{figure}[!t]
  \centering
  \includegraphics[width=160mm]{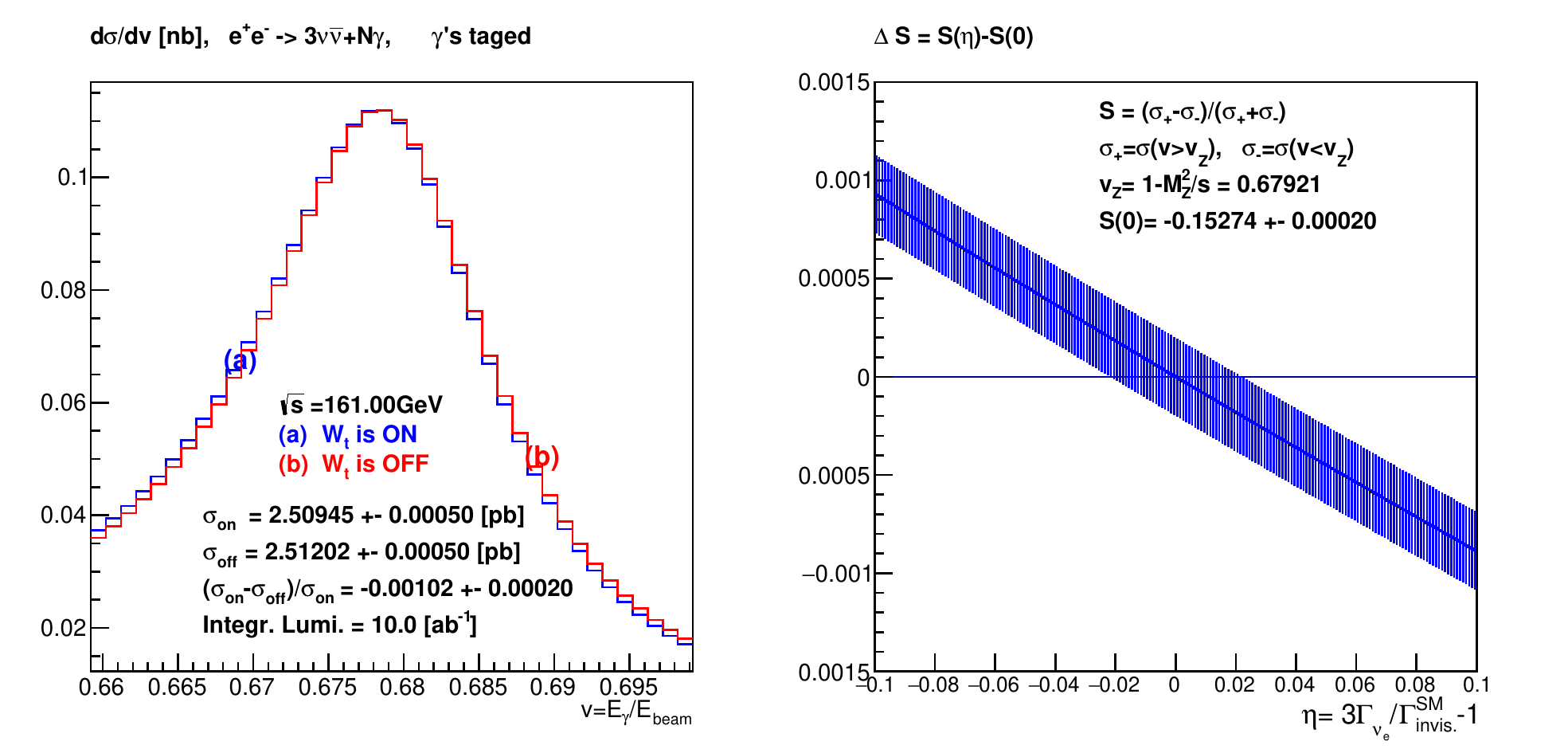}
  \includegraphics[width=160mm]{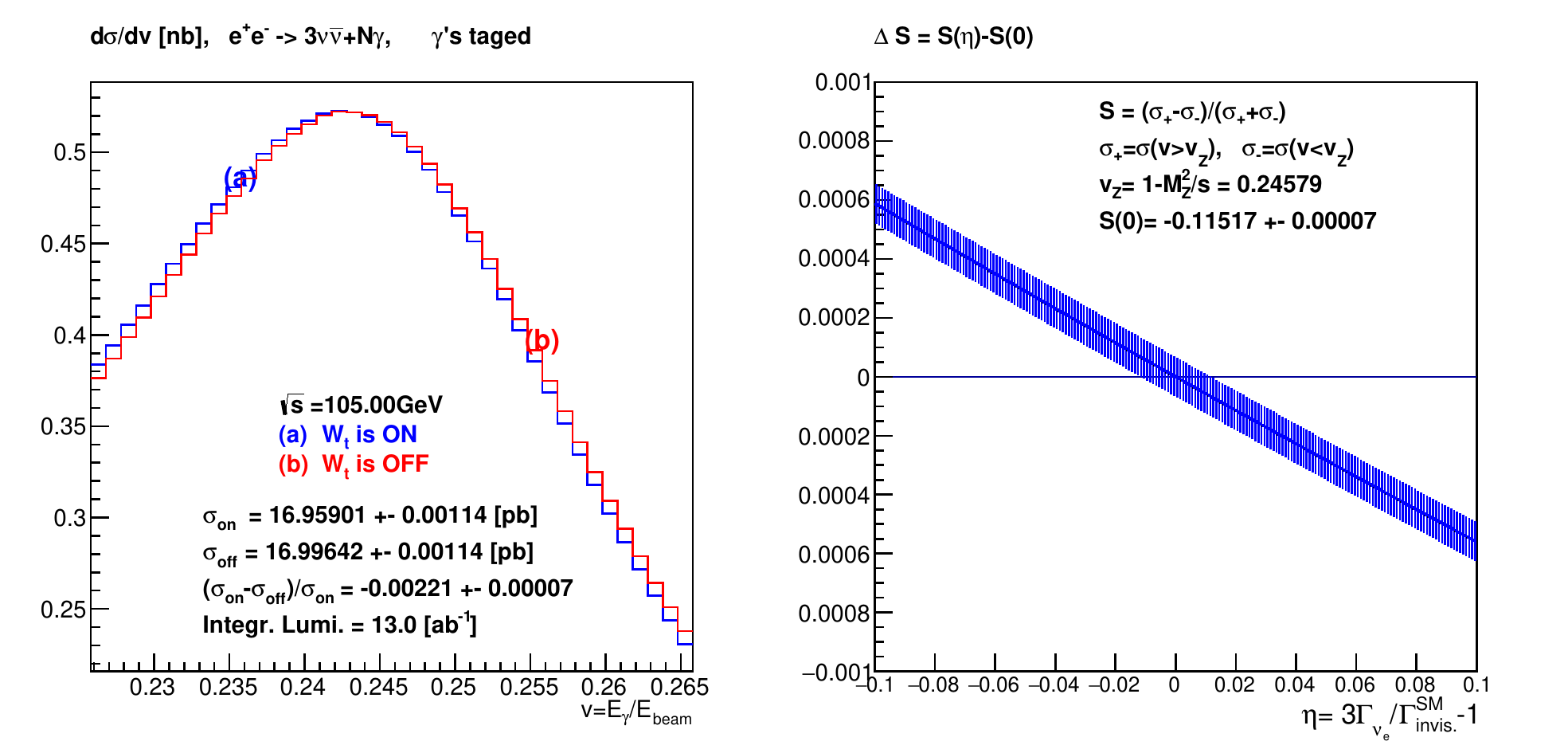}
  \caption{\sf
  Examining difference of the ZRR photon spectrum 
  for electron neutrino and muon neutrino channels
  due to $t$-channel W exchange.
  }
  \label{Fig:NuEle}
\end{figure}

Next, in Figure~\ref{Fig:NuDif2},
we examine ZRR photon energy spectrum near the $Z$ resonance,
focusing on the $W_t$ diagrams contribution, present in the electron neutrino case
and absent for two other SM neutrinos.
As compared to Born curve of Figure~\ref{Fig:BornNu}
the difference between electron and muon neutrino due to $Z_s\otimes W_t$ interference
is roughly of the same size but slightly diluted.
As expected, the $Z$ resonance shape is deformed due to ISR/QED,
increasing significantly the skewness of the resonance lineshape,
in fact more than $W_t\otimes Z_s$ interference effect.
Of course, ISR/QED effect can be subtracted using reliable Monte Carlo
calculation for the ZRR process like \kkmc.

Finally, in  Figure~\ref{Fig:NuEle} we examine how precisely one can
deduce $Z$ coupling to electron neutrino from the skewness of the ZRR spectrum
near $Z$ resonance for the anticipated FCC-ee integrated luminosity
of $10 ab^{-1}$ at 161GeV and $13ab^{-1}$ at 105GeV respectively%
\footnote{%
 These integrated luminosities correspond to 2 years operation with 2 detectors at 161GeV 
 and 6 months operation at 105GeV of FCC-ee.}.
The LHS plots of Figure~\ref{Fig:NuEle} shows ZRR photon spectrum for all three neutrinos
switching off and on $Z_s\otimes W_t$ interference, at two energies, 105GeV and 161GeV.

At $s^{1/2}=161$GeV
the ZRR cross section integrated over the range $v_Z\pm 0.02$ of the plot
is about 2.51pb.
With the 10ab$^{-1}$ integrated luminosity anticipated at FCC-ee,
can be measure with the $\pm5\cdot 10^{-4}$pb statistical error 
($\pm 2\cdot 10^{-4}$ relative error).
The entire $Z_s\otimes W_t$ interference effect in the integrated
ZRR cross section is $22\cdot 10^{-4}$ (relative) and is measurable.
However, in order to determine $g^{\nu_e}_Z$ from it with the $\sim 1\%$ precision 
we have to exploit the parameter
\begin{equation}
    S = \frac{ \sigma(v>v_Z) -\sigma(v<v_Z) }{\sigma(v>v_Z) +\sigma(v<v_Z)}
\end{equation}
parametrising the skewness of the $Z$ resonance curve.
In case of three neutrinos with equal SM couplings to $Z$, 
the prediction for the skewness parameter including QED effects and $Z_s\otimes W_t$ interference
is $S=-0.15274 \pm 0.00020$, with its statistical error 
adjusted for 10ab$^{-1}$ integrated luminosity%
\footnote{Statistical error in our MC runs was at least factor two smaller.}.
In order to see how sensitive is the skewness $S$ to $g^{\nu_e}_Z$ let us rescale 
$Z$ couplings to neutrinos as in Eq.~(\ref{eq:gZ}).

The corresponding change of $S(\eta)$ from the SM value $S(0)$ is presented%
\footnote{ In fact we rescale the differences between $\nu_e$ and $\nu_\mu$ 
  ZRR distributions in Figures.~\ref{Fig:NuDif2} by $(1+\eta)^{1/2}$.
  Near the $Z$ peak $W_t\otimes Z_s$ interference dominates this difference
  and $W$ contribution squared is negligible.
}
in the RHS plots of Figure~\ref{Fig:NuEle},
together with the statistical error band
for 10ab$^{-1}$ integrated luminosity.
From this figure it is easy to read the (statistical) precision of $g^{\nu_e}_Z$
at FCC-ee integrated luminosity is $\simeq 1\%$ at $s^{1/2}=161$GeV
and $\simeq 0.5\%$ at $s^{1/2}=105$GeV.
This is the main and very interesting result of the present study.

The reference value of $S(0)$ is coming entirely from the precision SM calculation,
hence it is important to know how much it is affected by
the perturbative SM higher order corrections.
In the above \kkmc\ calculation  the virtual \order{\alpha^1} EW corrections
for $s$-channel $Z_s$ and $\gamma_s$ exchanges
from DIZET library were always switched on.
In the additional MC run
we have checked that switching off  completely all \order{\alpha^1} 
virtual EW+QCD corrections in \kkmc\ 
(also vacuum polarization) causes shift of $S(0)$ by $0.0007$.
This is above the level of the experimental FCCee precision.
However, due to smallness of EW and QCD couplings,
we expect the size of unaccounted \order{\alpha^2} non-QED corrections
to $S(0)$ to be below $10^{-4}$ level.
We keep in mind that for $Z_s\otimes W_t$ interference (dominant in $S(\eta)$),
the present version of \kkmc\ features virtual EW+QCD corrections of DIZET
only for $Z_s$ but not yet for $W_t$.
Even for $Z_s$ exchange, in the presence of hard photon,
the EW virtual corrections of \kkmc\ are still incomplete%
\footnote{Strictly speaking they are extrapolated out from the soft photon regime.}.

In order to get an idea about the size of unaccounted higher order QED non-soft 
corrections in the SM prediction for $S(0)$
we have downgraded QED matrix element in \kkmc\
by one order, to exponentiated \order{\alpha^1} level,
i.e. to \order{\alpha^0} level for the ZRR process.
This has induced shift in $S(0)$ only by $4\cdot 10^{-4}$.
The unaccounted non-soft QED \order{\alpha^1} corrections to the ZRR process
are suppressed by factor $2\frac{\alpha}{\pi} \ln\frac{s}{m_e^2}\simeq 0.12 $,
hence we expect them to be safely below the FCC-ee experimental precision $\sim 10^{-4}$.

\begin{figure}[!t]
  \centering
  \includegraphics[width=0.69\textwidth]{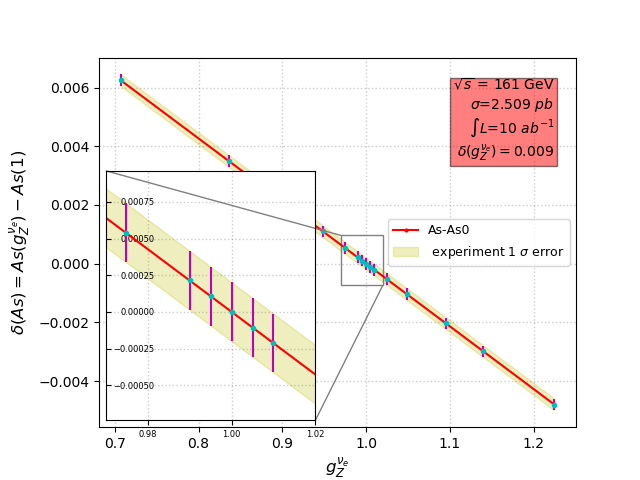}
  \includegraphics[width=0.69\textwidth]{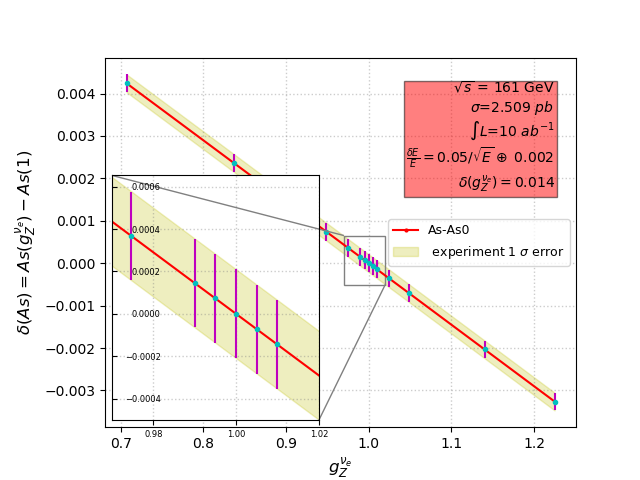}
  \caption{\label{fig:resolution} Experimental sensitivity of $g_Z^{\nu_e}$. 
  Detector resolution effect off (left plot) and on (right plot). 
  A 50$\%$ degradation is observed but the sensitivity still remains very good.
  }
\end{figure}

\section{Detector resolution effect}

Let us now examine the effect of the detector resolution. To this end, we assume homogeneous calorimeter with a reasonably good energy resolution of :

\vskip 10pt
\begin{equation}
\begin{array}{lcllcllc} 
\frac{\sigma (E_\gamma)}{E_\gamma } & = &  \frac{0.05}{\sqrt{E_\gamma} } \oplus 0.002
\end{array}
\label{eq:e_resolution}
\end{equation}

\noindent The energy resolution of photons can be extracted experimentally from the reconstructed $\mu\mu\gamma$ events, for which the muons reconstructed angles are very precise. Thus one can achieve a 1C fit of such events and obtain a precise $E_\gamma$ measurement, which is then compared to the energy measured with the calorimeter. This allows one to verify that the energy scale is correct. With the resolution in equation~(\ref{eq:e_resolution}), one observes as expected a degradation of the sensitivity on $g_Z^{\nu_e}$. However the sensitivity remains excellent at the level of 1.4 $\%$ as can be seen in Figure~\ref{fig:resolution}. With this error for $g_Z^{\nu_e}$ and assuming that $N_\nu$ will be measured with negligible error at FCC-ee and that there is no invisible new particles coupled to $Z$, one gets a sensitivity of $4.8\%$ on $g_Z^{\nu_\tau}$ from equation~\ref{eq:N_nu} (the error on $g_Z^{\nu_\mu}$ from PDG was used). 

Should the stochastic term be twice as worse (i.e. $0.1/\sqrt{E}$, which is typical for a sampling calorimeter), the sensitivity on $g_Z^{\nu_e}$ would be $2.4\%$. This "rapid" degradation implies that an excellent knowledge on the calibration of $E_\gamma$ is crucial. Obviously a very good calorimeter energy resolution is very important for this measurement.

\section{Summary and outlook}
Main result of our study is that by means of exploiting interference between $s$-channel $Z$ exchange
and $t$-channel $W$ exchange in the $e^+e^-\to X+\gamma$ process,
the coupling constant of $Z$ boson to electron neutrino and $Z$ 
can be measured at the FCC-ee high luminosity
with the statistical error of $\sim 1\%$, 
that is a factor $\sim 20$ better than the present error.
The above result is obtained assuming observation of at least one photon with an angular distance
from both beams above $15^\circ$ and the energy above 10\% of the beam energy.
This important and encouraging result was obtained assuming integrated luminosity 10 and 13ab$^{-1}$
for centre of the mass energy 161 and 105GeV.
Keeping systematic error due to calorimeter energy resolution at the similar $1\%$ level
is within the reach of the available detector technology.
It should be stressed that the above high precision requires comparison of experimental
data with high quality Monte Carlo even generator, because multiphoton QED effects
have to be removed in order to pin down the valuable $Z_s\otimes W_t$ interference effect.

The presented study is still preliminary and there is a number of issues requiring further studies:
(1) virtual corrections for $W_t$ contribution in \kkmc\ matrix element has to be completed,
(2) the size and shape of the QED deformation of the $Z$ peak in ZRR 
obtained from \kkmc\ should be cross-checked using independent calculation
(3) calculating complete \Order{\alpha^2} EW loop corrections 
to $e^+e^-\to \nu\bar\nu$ process
(\Order{\alpha^1} to ZRR $e^+e^-\to \nu\bar\nu\gamma$ subprocess) 
should be seriously considered,
(4) dominant \Order{\alpha^3} QED non-soft corrections (in our convention)
should be estimated/calculated.

There are also several other improvements in the analysis front, which needs to be studied, 
such as carrying a full fit of the $v$ spectrum instead of measuring its asymmetry 
and/or optimizing the $v$ range. 
Also, as already mentioned, study of the interference effect 
at low and high $v$ range%
\footnote{This would require introducing rescaling $\eta$ parameter directly in the \kkmc\
  matrix element.}
might be useful to improve the sensitivity on $g_Z^{\nu_e}$.

\vspace{10mm}
\noindent
{\bf\large Acknowledgments}

We would like to thank M.~Skrzypek and Z.~W\c{a}s for useful discussions.

\newpage

\end{document}